# Serendipitous 2MASS Discoveries Near the Galactic Plane: A Spiral Galaxy and Two Globular Clusters

Robert L. Hurt, Tom H. Jarrett, J. Davy Kirkpatrick, Roc M. Cutri

Infrared Processing & Analysis Center, MS 100-22, California Institute of Technology,
Jet Propulsion Laboratory, Pasadena, CA 91125
hurt@ipac.caltech.edu, jarrett@ipac.caltech.edu, davy@ipac.caltech.edu, roc@ipac.caltech.edu

Stephen E. Schneider, Mike Skrutskie

Astronomy Program, University of Massachusetts, Amherst, MA 01003
schneider@astro.umass.edu, skrutski@astro.umass.edu

Willem van Driel

United Scientifique Nançay, Obs. de Paris–Meudon, Meudon Cedex, CA 92195, France
vandriel@obspm.fr

## Abstract

We present the basic properties of three objects near the Galactic Plane—a large galaxy and two candidate globular clusters—discovered in the Two Micron All Sky Survey (2MASS) dataset. All were noted during spot-checks of the data during 2MASS quality assurance reviews. The galaxy is a late-type spiral galaxy (Sc–Sd), ~11 Mpc distant, at l = 236.82°, b = -1.86°. From its observed angular extent of 6.3' in the near infrared, we estimate an extinction-corrected optical diameter of ~9.5', making it larger than most Messier galaxies. The candidate globular clusters are ~2–3' in extent and are hidden optically behind foreground extinctions of $A_v$ ~18–21 mag at l ~ 10°, b ~ 0°. These chance discoveries were not the result of any kind of systematic search but they do hint at the wealth of obscured sources of all kinds, many previously unknown, that are in the 2MASS dataset.

Key words: galaxies: photometry—galaxies: spiral—(Galaxy:) globular clusters: general—surveys—infrared radiation

# 1. Introduction

The Milky Way Galaxy is both a source for and an obstacle to astronomical study. The dust found in the plane of the Milky Way forms a veil that obscures our view of more distant sources at visible wavelengths. In extragalactic astronomy the region near the Galactic Plane suffering from this obscuration has been dubbed the Zone of Avoidance (ZoA) due to the systematic incompleteness of galaxy catalogs here. The high extinctions in the ZoA also affect Galactic astronomy as well, since the most distant objects within the Milky Way are as severely extincted as extragalactic sources.

The Galactic dust clouds are less of an obstacle at longer wavelengths, becoming increasingly transparent in the infrared and radio. Extensive surveys in the mid-infrared through radio regimes have uncovered a wealth of sources hidden to optical astronomy within the Milky Way and beyond. For example, dust-rich galaxies brighter than 1–2 Jy at 60 µm are evident in the IRAS dataset at virtually all Galactic latitudes while hydrogen-rich galaxies within the ZoA have been identified by 21-cm surveys (e.g. the Dwingeloo Obscured Galaxies Survey and the Parkes multibeam survey; Henning et al. 1998; Henning et al. 2000). Such surveys trace a variety of emission processes including warm dust, atomic/molecular gas content, and thermal/nonthermal electron radiation, but are not specifically sensitive to emissions of stellar photospheres. In extragalactic astronomy these limitations select against inactive and gas-poor galaxies (including ellipticals), while in Galactic work, multitudes of stars and clusters are hidden behind the dust.

An effective way to penetrate the ZoA to detect obscured galaxies, star clusters, and other sources dominated by stellar photospheric emission is to utilize surveys in the near infrared. Extinctions at K band are about 1/10th those in the optical so it is possible to probe much deeper into the ZoA by using infrared wavelengths. The advent of large infrared arrays and increased data processing capabilities have made possible sensitive near infrared surveys that have high angular resolution and can cover large areas, as with the Deep Near-Infrared Southern Sky Survey (DENIS; Epchtein 1998), or the entire sky, as with the Two Micron All Sky Survey (2MASS; Skrutskie et al. 1997).

2MASS is a ground-based all-sky survey that is imaging the entire sky in three near infrared bands at J (1.25 µm), H (1.65 µm), and $K_s$ (2.17 µm). Survey data are reduced in an automated pipeline (Cutri et al. 2000) that produces astrometrically and photometrically calibrated Atlas Images (512x1024 pixels at 1"/pixel) in the three survey passbands. It also extracts catalogs of point and extended sources including positions and fluxes. The 10:1 signal-to-noise limits for the point and extended source catalogs are J ≤ 15.8, H ≤ 15.1, and $K_s$ ≤ 14.3 mag and J ≤ 15.0, H ≤ 14.2, and $K_s$ ≤ 13.5 mag, respectively. The final catalog is anticipated to include ~300 million stars and ~2 million galaxies, a significant fraction of which will represent previously unseen and unknown sources. To date several hundred ZoA galaxies have been identified in a systematic search of the region covered by the first incremental release of 2MASS data (Jarrett et al, 2000a), and another large spiral galaxy has been found during data processing (Howard, 2000). By the end of the survey, many thousands of new ZoA galaxies, as well as multitudes of obscured sources within our own Galaxy, should appear in the final catalog.

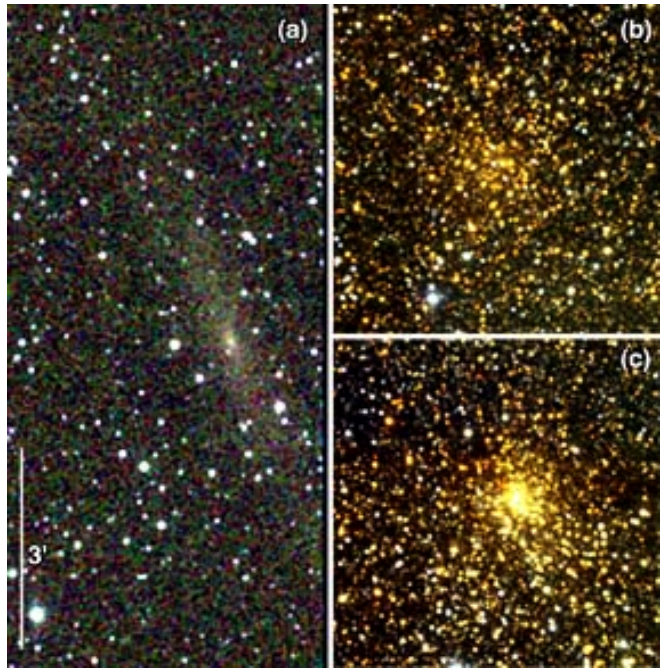

Fig. 1.—Near Infrared Color Composite of a galaxy and two globulars. The 2MASS galaxy 2MASXI J0730080-220105 is shown in panel (a) while 2MASS-GC01 and 2MASS-GC02 are in panels (b) and (c). The J, H, & $K_s$ bands have been mapped onto blue, red, and green channels, respectively. Image resolutions are 1"/pixel and all three images are at the same size scale. A scale reference bar, 3' in length, appears in panel (a).

We present in this paper three intriguing objects discovered in the 2MASS dataset: a large spiral galaxy and two candidate globular clusters (see Fig. 1). The new galaxy lies in a region with Galactic V-band extinctions of $A_V$ ~ 4 mag (see Table 1) and

## Table 1
### Galactic Extinction Estimates

| Source | Extinctions | | | |
|---|---|---|---|---|
| | V | J | H | $K_s$ |
| 2MASXI J0730080-220105 | | | | |
|    IRAS/COBE[a] | 4.9 | 1.3 | 0.81 | 0.49 |
|    HI[b] | 4.1 | 1.1 | 0.67 | 0.41 |
| GC01 | | | | |
|    IRAS/COBE[a] | 71.6 | 19.6 | 11.7 | 7.2 |
|    2MASS | 20±2 | 5.0±0.5 | 3.0±0.3 | 1.8±0.2 |
| GC02 | | | | |
|    IRAS/COBE[a] | 37.6 | 10.3 | 6.2 | 3.8 |
|    2MASS | 18±2 | 4.5±0.5 | 2.7±0.3 | 1.7±0.2 |

[a]Values derived from extinction maps from combined IRAS and DIRBE datasets (Schlegel, Finkbeiner, & Davis 1998). Extinctions are for entire line-of-sight through the Galactic Plane and are thus upper limits to the foreground extinctions for GC01 & GC02.

[b]Values derived from HI column densities and HI/dust ratios (Hartmann & Burton 1997; Bohlin, Savage, & Drake 1978). HI estimates are unreliable for GC01/GC02 region due to optical depth saturation through dense columns close to the Galactic Center (derived $A_v$'s are only 6–8 mag).

was independently discovered in the HI Parkes Shallow Survey of the Southern Zone of Avoidance (Henning, et al. 2000). The globulars are in regions of much higher extinction ($A_v \sim 20$ mag) and have not appeared in previously-published catalogs. Each object of interest was noted by eye during visual spot-checks of a small subset of Atlas Images (Cutri et al. 2000) chosen from each night's production run. Such quality checks of 2MASS production results are used to help identify possible problems in the data reliability (clouds, electronic glitches, airplane trails, etc.), but occasionally uncover objects of scientific interest as well.

## 2. The New Galaxy

The new galaxy presented here is noteworthy due both to its proximity and large angular size. It was first identified in 2MASS data in the fall of 1998 during pipeline processing (Cutri et al. 2000) of Northern Hemisphere observations from the night of 3 May, 1998. The source was automatically extracted by the extended source subsystem (Jarrett et al. 2000b) but was recognized as particularly significant during a quality spot-check of because of its obvious extent of several arcminutes and morphology consistent with a late-type spiral galaxy. It appears in the extended source catalog for the Second Incremental Data Release as source 2MASXI J0730080-220105, following established 2MASS naming convention. For convenience in this paper we will employ a shortened version of the official designation: 2MX 0730-2201.

This new galaxy was previously unknown due to the foreground extinction from the ZoA. A check of both the NASA/IPAC Extragalactic Database (NED)[1] and SIMBAD[2] at the time of discovery turned up no published references to this source. Prior optical discovery was unlikely since Digitized Sky Survey plate for this region shows only a very faint nebulosity that would be easily overlooked in any but the most extreme image stretches. This galaxy has been independently identified and published as part of the HI Parkes Shallow Survey as HIZSS 012 (Henning et al. 2000). The only other extragalactic sources found in NED within 1° of 2MX 0730-2201 are CGMW 2-0931 (12.4' away) and ZOAG G237.32-01.28 (45.3' away), both of which are near small (diameter < 30"), extended 2MASS detections. No other potential companion extragalactic sources of significant size within a 1° radius are noted in the 2MASS Extended Source Catalog.

### 2.1. 2MASS Photometric Properties

Near infrared properties for this newly discovered galaxy are derived from analysis of the Atlas Images in this highly crowded field (see Fig. 1a). 2MASS pipeline processing produces mutually registered J, H, $K_s$ Atlas Images, about 8.5' wide, interpolated onto a 1"/pixel grid. The galaxy nucleus is about 1.4' away from the edge of this Atlas Image and its southwestern disk extends onto the overlapping edge of an adjacent Atlas Image from a subsequent night. Since the full galaxy does appear on the discovery image we have not attempted to mosaic the images for analysis purposes.

---

[1] The NASA/IPAC Extragalactic Database (NED) is operated by the Jet Propulsion Laboratory, California Institute of Technology, under contract with the National Aeronautics and Space Administration.

[2] SIMBAD is operated by the Centre de Donnés astronomiques de Strasbourg and support for use by US astronomers is provided by the National Aeronautics and Space Administration.

To allow our flux measurements to push down into the noise, we subtracted out a polynomial fit to the sky background in all bands. A 2D cubic fit was used in all bands, with stars and the region around the galaxy masked out of the fit; fitting was accomplished using procedures written in IDL. The background model has peak-to-peak variations of 0.7, 2.1, & 0.7 DN/pixel (or counts/pixel) against backgrounds around 276, 778, & 620 DN/pixel in the J, H, & $K_s$ bands (source surface brightness in the extended disk is less than 1 DN/pixel). The background subtraction was necessary since rapid variations in atmospheric OH airglow emission can introduce structure into the Atlas Image backgrounds, strongest at H band, at levels exceeding those seen in the faint extended emission in galaxies.

We estimate the Hubble type of the galaxy to be Scd–Sd based on its near infrared morphology. In the image (Fig. 1a) we see a tiny nucleus and extended faint disk lacking any obvious spiral structure. The galaxy is inclined significantly to the line-of-sight. It is difficult to reliably map observed infrared morphologies of galaxies onto optical Hubble Types since the stellar populations traced by the two wavelength regimes are rather different. Critical optical features on which such classifications are based (e.g. spiral arm shapes that stand out because of strong dust lanes and localized populations of young, hot stars) can be difficult to determine in the near infrared due to reduced contrast. Our Hubble type determination is based on anecdotal examination of a large number of 2MASS spiral galaxies with similar angular sizes, inclinations, and morphologies. We find that later-type galaxies tend to have lower surface brightnesses and nuclei that are significantly less prominent than those seen in early-type galaxies. We conclude that the new galaxy is no earlier than Sc (where brighter, distinct nuclei are more apparent) but not as late as Sm (where the nuclei are not distinct from the disk and the near infrared surface brightnesses are very low, occasionally below 2MASS detection limits). A more accurate classification will be possible as better statistical studies of infrared galaxy properties from 2MASS and other surveys become available.

The galaxy position and source geometry were determined from analysis of an image sum of all three bands to maximize signal-to-noise of the faint extended structure (see summary in Table 2). The central position of the galaxy, 7h30m08.1s, -22° 01' 06" (J2000), is taken from the extended source database entry and matches independently-derived estimates from the Atlas Images to better than 1". The position angle of the extended disk, 23±3° (E of N), was determined using the "ellipse" fitting routine in IRAF, with greater weight given to the ellipses fit to the outer disk. Further analysis was performed using an image where resolved stars were masked. The inclination was determined by computing average brightnesses in elliptical annuli centered on the galaxy nucleus and minimizing the RMS deviations of these averages. An inclination of 70±5°, was found to fit best the outer annuli (an infinitely thin disk was assumed; corrections for finite disk thickness are inconsequential within our errors). The more spherical bulge component of the nucleus violates the assumption of a thin disk for this simple calculation and tends to bias the inner regions to smaller inferred inclinations.

**Table 2**
**Galaxy Properties**

| Property | Value |
|---|---|
| 2MASS Designation | 2MASXI J0730080-220105 |
| Position | |
|     RA (J2000) | $7^h30^m08.1^s$ |
|     Dec (J2000) | -22° 01' 06" |
|     l | 236.8173 |
|     b | -1.8509 |
| Inclination | 70±5° |
| Position Angle | 23±3° |
| Hubble Type | Scd–Sd |
| Magnitudes[a] | |
|     J | 10.3 (9.1) ± 0.1 mag |
|     H | 9.5 (8.7) ± 0.1 mag |
|     $K_s$ | 8.9 (8.5) ± 0.1 mag |
| Diameter | |
|     Observed NIR | 6.3' |
|     Inferred optical[b] | 9.5 ± 1.5' |
| Distance[c] | 11 Mpc |
| HI Properties | |
|     HI flux | 76±8 Jy km/s |
|     $V_{hel}$ | 779±1 km/s |
|     $V_{corr}$ | 803±1 km/s |
|     Linewidth $W_{20,c}$ | 302±2 km/s |
|     $M_{HI}$ | $2.2 \times 10^9 M_{sun}$ |

[a] Observed magnitudes are listed first followed by extinction-corrected magnitudes in parentheses. Extinction corrections are the average of the IRAS/COBE and HI extinction estimates from Table 1.

[b] Estimate for the B-band 25 mag/arcsec$^2$ isophote.

[c] Average of values derived from HI velocity/Hubble flow and H-band Tully-Fischer relationships.

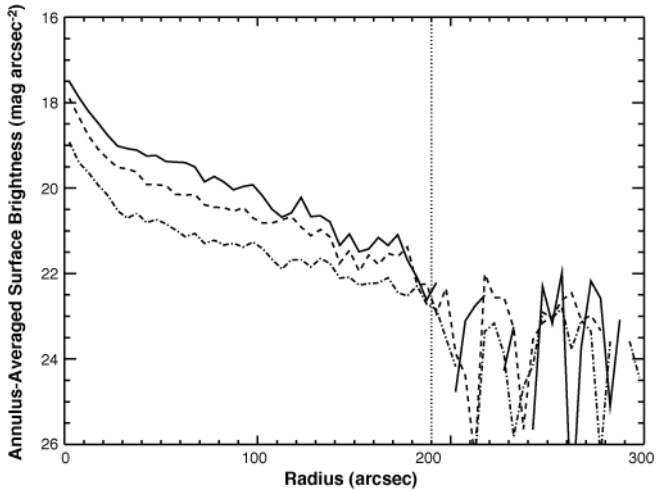

Fig. 2.—Radial Brightness Profiles for 2MASXI J0730080-220105. Annulus-averaged radial brightness profiles are shown for J, H, & $K_s$ bands. An inclination of 70° and position angle of 23° were assumed (see text). The averages drop into the noise in all bands at major axis radii > 190"; this adopted infrared radius is marked with a vertical dashed line.

Surface brightnesses and integrated magnitudes were measured in each band adopting these geometrical factors. The radial brightness profiles (Fig. 2) are calculated from averages within elliptical annuli in the masked images. The diffuse disk emission can be distinguished in J band out to the ~22 mag/arcsec$^2$ isophote and in H & $K_S$ to the ~21 mag/arcsec$^2$ isophote, after which the radial averages clearly drop into the noise. This isophote is at a radius of 190 ± 15" in all bands, corresponding to an observed infrared major-axis diameter of $D_{IR}$ = 6.3 ± 0.3'. Integrated magnitudes (see Table 2) were calculated from curves-of-growth derived from the average radial brightness profiles and are statistically reliable to better than 0.1 mag.

The corresponding unextincted optical diameter of 2MX 0730-2201 is likely to be significantly larger than the observed infrared diameter of 6.3'. During anecdotal studies of galaxies during quality review, we have taken note of late type galaxies with similar inclination and disk/bulge morphologies. In this sample of several dozen sources we note that the full observed 2MASS extent always fall short of the RC3 optical diameters ($D_{25}$ measured to the B band 25 mag/arcsec$^2$ isophote; de Vaucouleurs, et al. 1991), typically by factors of 0.73 ± 0.1. The full extent from eyeball estimates is roughly equivalent to the $D_{IR}$ diameter cited above, though it varies with sensitivity of the Atlas Image and is affected by foreground extinction. Applying extinction corrections to the radial brightness profiles and extrapolating them to the noise limit we conclude the observed $D_{IR}$ falls short of the extinction-corrected diameter by a factor of 0.8–0.9. Using a conservative extinction correction of 0.9 and our anecdotal infrared/optical ratios, we derive a $D_{25}$ estimate of 8.4–11.0'.

In a more systematic study, Jarrett and Corwin (2000) report that the ratio of the near infrared diameter $D_{20}$ (measured to the K band 20 mag/arcsec$^2$ isophote) to the optical diameter $D_{25}$ is systematically smaller than unity for RC3 galaxies detected by 2MASS. The ratio is smallest for late-type, unbarred, Sc-Sd galaxies, with $D_{20}/D_{25}$ = 0.56 ± 0.17. Correcting for $A_K$ = 0.45 (Table 1), the new galaxy has a corresponding diameter of $D_{20}$ = 4.3 ± 0.2', leading to an optical diameter estimate of $D_{25}$ = 5.2–11.0'. The upper limit to this estimate is consistent with our previous estimate though we feel the lower end of this range is unrealistic since it falls well short of the observed near infrared $D_{22}$ of this galaxy. The $D_{20}/D_{25}$ statistics do not include late type galaxies of diminished surface brightness for which 2MASS $D_{20}$ data are not compiled and may be biased towards slightly larger values of this ratio, leading to a slight underestimate of $D_{25}$. Taking the overlap between our two diameter estimates we arrive at a likely extinction-corrected optical diameter of $D_{25}$ = 9.5 ± 1.5'. This is about as large as M51 and larger than most of the spiral galaxies in the Messier catalog. As such, this is the largest diameter galaxy independently discovered in the near infrared since the advent of infrared array detectors. The previous record-holders are Maffei 1 & 2, first noted in infrared-sensitive photographic plates of the Galactic Plane (Maffei 1969).

### 2.2. HI Mass, Distance, and Luminosity

The galaxy 2MX 0730-2201 was observed in the 21 cm HI line with the Nançay radio telescope in an attempt to detect the neutral gas expected for a spiral galaxy. The Nançay telescope (see, e.g., Theureau et al. 1998) is a meridian transit-type instrument with an effective collecting area of 7000 m$^2$ (equivalent to a 94-m diameter parabolic dish). Due to the elongated geometry of the mirrors, it has a HPBW of 3.6' × 22' (α × δ) at 21 cm wavelength. Given the dimensions and orientation of the galaxy, we expect to have detected its entire HI content in a single pointing with the telescope. Typical system temperatures were ~40 K.

The observations were made in November and December 1998, using a total of about 7 hours of telescope time over a period of 9 days. We observed in total power (position-switching) mode using consecutive pairs of two-minute on- and two-minute off-source integrations. Off-source integrations were taken approximately 30' E of the target position. The autocorrelator was divided into two orthogonal linear polarizations, each with 512 channels and a 6.4 MHz bandpass, providing a velocity coverage of about 1200 km/s and a velocity resolution of about 3.1 km/s. At first, the center frequencies of the two banks were tuned to a radial velocity of 250 km/s, in order not to miss possible HI emission at negative velocities. After the detection of the strong line signal, the center frequencies were shifted to 750 km/s.

We reduced our HI spectra using the standard DAC and SIR spectral line reduction packages available at Nançay. We subtracted a third order polynomial baseline, averaged the two polarizations, and applied a standard declination-dependent conversion factor to convert from units of $T_{sys}$ to flux density in mJy, following the procedure given in Matthews, Gallagher, & van Driel (1998) and Matthews & van Driel (2000).

The reduced Nançay spectrum (Figure 3) shows the typical double-horned profile of an inclined spiral disk. The center heliocentric velocity, determined from the average of the 50% points, is 779±1 km/s. The profile widths at the 50% and 20% levels of the peak flux density are 272±1 and 284±2 km/s, respectively and the integrated line flux is 76±8 Jy km/s. Errors in the radial velocity and velocity widths were estimated following the procedure given in Matthews et al. (1998), which is based on Fouqué et al. (1990). These HI measurements are consistent with the values and associated uncertainties given in Parkes HI multibeam observations of this source (Henning et al. 2000).

The galaxy's distance as estimated from its redshift is potentially affected by large scale flows and local dynamics. The POTENT model of Dekel, Bertschinger, and Faber (1990) indicates that the peculiar velocity due to streaming at this position is only -24 km/s, yielding a corrected redshift of 803 km/s. There are few galaxies with accurate distances in the ZoA, so this estimate is extrapolated from galaxies at fairly large angular separations, but there is no strong indication that a large offset from the observed redshift is present. Assuming a Hubble constant of 75 km/s Mpc$^{-1}$ yields a distance of 10.7 Mpc.

We can also estimate the luminosity and distance with the Tully-Fisher (TF) relation. Based on preliminary analyses, the Hubble Space Telescope Key-Project TF relation (Sakai et al. 1999) is consistent with the TF relation found using 2MASS H-band magnitudes and Arecibo HI line data measured at 50% of peak flux density. Adopting an inclination of 70°, the corrected $W_{20,c}$ line width is 302 km/s. This corresponds to an absolute magnitude of $H^c_{-0.5}$ = -21.52, where $H^c_{-0.5}$ is the aperture magnitude extrapolated to the radius equivalent to half the blue light of the galaxy (from Aaronson et al. 1982).

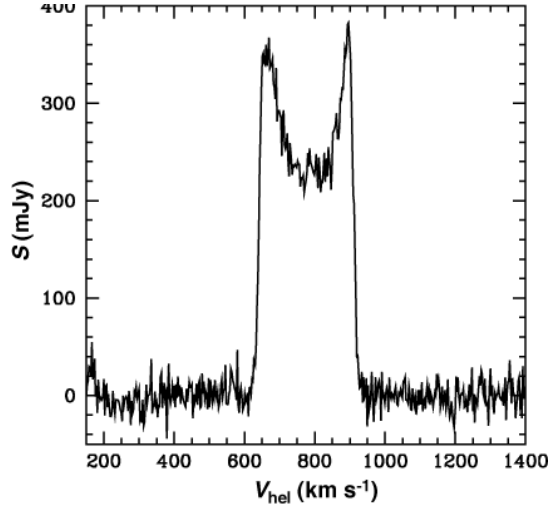

Fig. 3.—HI Spectrum for 2MASXI J0730080-220105. This Nançay 21 cm spectrum shows the classic double-horned signature of a spiral disk galaxy. Its integrated flux is 76±8 Jy km/s and its heliocentric velocity is $V_{hel}$ = 779 ± 1 km/s.

To estimate the Galactic dust extinction we use two empirical estimates inferred from COBE/IRAS far-infrared emissions (Schlegel, Finkbeiner, & Davis 1998) and HI column densities (Hartmann & Burton 1997; Bohlin, Savage, & Drake 1978). Correcting for $A_H$ = 0.74 mag (the average of the HI and COBE/IRAS method estimates), $H^c$ = 8.71 as an approximation of $H^c_{-0.5}$, yields an estimated distance of 11.1 Mpc. Since the galaxy is nearly edge-on, the extinction correction might be slightly larger, implying a somewhat smaller distance. The new galaxy is therefore well outside the Local Group but is somewhat closer than the Virgo cluster (at 16 Mpc; Graham et al. 1999).

Adopting a distance of 11 Mpc, the HI mass of 2MX 0730-2201 is 2.2 x 10$^9$ M$_{sun}$, showing it is a fairly gas-rich system. The H-band luminosity is 9.6 x 10$^9$ L$_{sun}$ for an assumed absolute solar magnitude of 3.44 in the H-band (Bothun 1984). This yields an HI mass-to-light ratio of $M_{HI}/L_H$ = 0.23 M$_{sun}$/L$_{sun}$. According to the correlations between HI and H-band luminosities found by Bothun (1984), this ratio

## 3. Two New Globular Clusters

A new pair of candidate globular clusters has also been identified in 2MASS data (Figs 1b & c). These clusters are both at very low Galactic latitudes ($|b| < 0.7°$) and are separated by just over 1° (see Table 3). Both of these clusters were noted by eye during the quality review of arbitrarily selected Atlas Images, though the second was first noted as a probable globular by 2MASS telescope operator Joselino Vasquez in his log. Photometry for the brightest stars in these clusters was extracted by the pipeline and a portion of the first cluster was flagged as an extended source candidate at H & $K_s$ bands (offset ~ 20" from the cluster center). Since the one extended source identification did not meet the source selection criteria for the Second Incremental Release (it was too close to the heavily confused Galactic Center region), there are no standard 2MASS designations appropriate for these ensemble sources. We therefore will dub them 2MASS-GC01 and 2MASS-GC02, hereafter shortened to GC01 and GC02 for this paper.

The globular cluster candidates GC01 & GC02 appear particularly striking in three-color images because of their yellow-red appearance (Figs 1b & c). These colors, dominated by H & $K_s$ bands, indicate the high extinctions found only ~10° from the Galactic Center. COBE/IRAS dust models predict a total visual extinction of ~72 mag towards GC01 and ~38 mag towards GC02. These are total extinctions of dust columns extending through the entire Milky Way disk and thus represent upper limits to the foreground extinctions of these clusters since they lie somewhere in the middle of these dust columns. Dust extinctions derived from HI column densities are much lower, only ~6–8 mag, and suffer from optical depth saturation through such dense regions.

A more reliable estimate of the foreground extinctions can be derived from the photometric colors of these clusters. Globular clusters typically have relatively tight J-$K_s$ and H-$K_s$ color distributions, dominated by similar populations of giant branch stars, so it is possible to estimate foreground extinctions using color-color diagrams. Such diagrams for GC01 and GC02 are presented in Figures 4a & 4b. Three populations of stars are plotted, all drawn from pipeline-extracted 2MASS photometry: large dots are stars from within a box tightly bounding the observed extent of the cluster; small dots are field stars near (but not including) the clusters; crosses are stars belonging to NGC 6496 (chosen as a fiducial, relatively unextincted globular, ~10° above the Galactic Plane). The field stars show typical scatter along the reddening vector while the cluster members (which have significant scatter from foreground source contamination and photometry errors induced by source confusion) clump towards the higher extinctions.

We have also calculated integrated ensemble colors for the inner cores of these clusters. This is a semi-independent check on the colors since point source photometry exists only for the brightest unconfused stars in the periphery of these clusters. Core magnitudes, using ~1–1.5' diameter apertures, include the flux contribution of the confused core stars for which point source photometry is not available. Uncertainties were estimated from the variation in colors seen in different choices of aperture. These integrated core colors are shown as ellipses (approximating 1 sigma errors) in Figure 4 and are consistent with the centroids of the individual stars. We note that such core magnitudes can be strongly dominated by a handful of the brightest stars, but we have attempted to select apertures excluding such obvious bright sources and are more dominated by the cluster diffuse, confused emission.

The photometrically-derived extinctions are large, consistent with the IRAS/COBE upper-limit estimates for this region. GC01 has a visual extinction of $A_v = 21.5 \pm 1.0$ mag, in a region where field stars show extinctions as high as $A_v$ ~30 mag. GC02 is slightly less extincted, with $A_v = 18.0 \pm 1.0$ mag, with field stars reaching up to $A_v$ ~ 25 mag. Such large extinctions preclude any likely detection by optical means; there is no trace of these clusters

**Table 3: Cluster Properties**

| Property | GC01 | GC02 |
|---|---|---|
| Position | | |
|   RA (J2000) | $18^h\ 08^m\ 21.81^s$ | $18^h\ 09^m\ 36.5^s$ |
|   Dec (J2000) | -19° 49' 47" | -20° 46' 44" |
|   l | 10.4717 | 9.7821 |
|   b | +0.098 | -0.6151 |
| Estimated $K_s$ Diameters[a] | 3.3±0.2' | 1.9±0.2' |
| Integrated Core Magnitudes[b] | | |
|   J | 12±2 | 9.5±0.4 |
|   H | 8.6±0.4 | 6.6±0.2 |
|   K | 7.2±0.4 | 5.5±0.2 |

[a] Determined from inspection of 2MASS $K_s$ Atlas Images.
[b] Derived from apertures encompassing brightest H & $K_s$ emission in center of cluster; errors estimated from variations seen between slightly different aperture choices.

is quite typical of Sc galaxies.

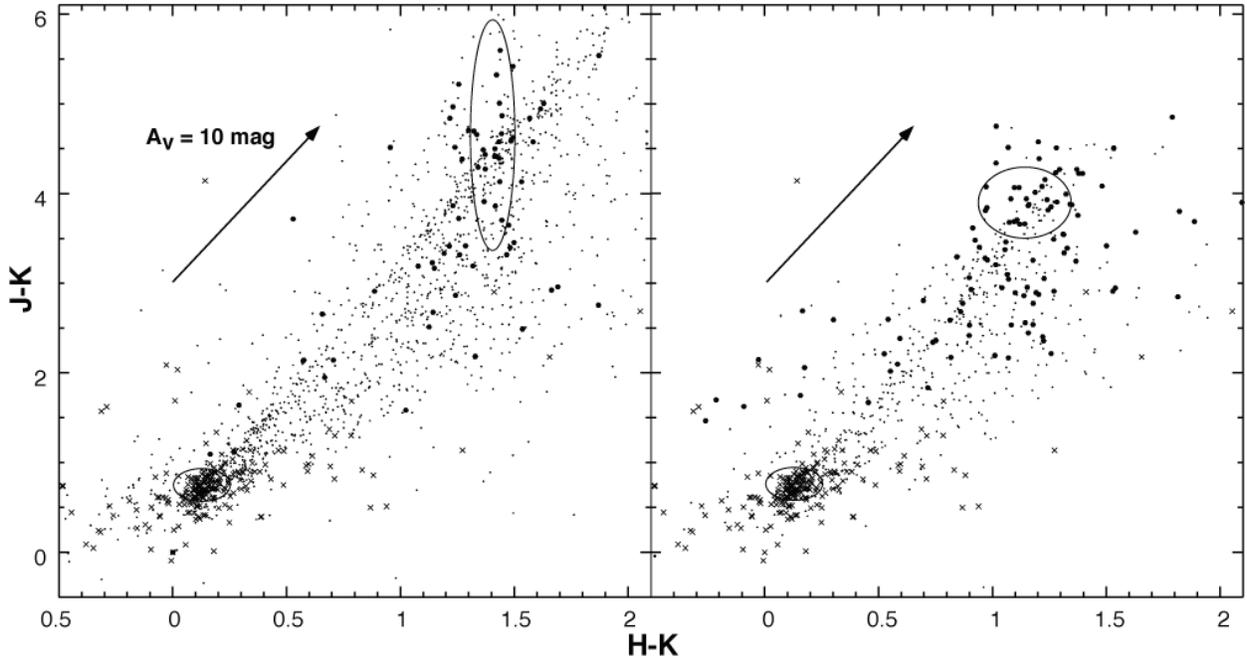

Fig. 4.—Globular Cluster Color–Color Plot. Near infrared J-$K_s$ vs. H-$K_s$ plots for GC01 (a) and GC02 (b) are shown with reddening vectors for $A_v$ = 10 mag. Stars associated with the clusters are plotted with large circles. Field stars from areas near the clusters are plotted with smaller dots. Stars from a reference globular cluster NGC 6496 are plotted with crosses (representing an unextincted sample). Approximate error ellipses for integrated core-magnitude colors are shown for GC01, GC02, & NGC 6496 (GC01 is particularly faint at J and has a correspondingly larger error in the Y direction). From the offset of the cluster core stars along the reddening vector from the reference cluster we estimate foreground extinctions of $A_v$ ~ 20 mag for GC01 and $A_v$ ~ 18 mag for GC02.

in Digitized Sky Survey images of this region.

Why are GC01 & GC02 likely to be globular clusters rather than open clusters? Their morphologies are certainly consistent with those seen in other globulars, highly concentrated towards their centers. However, there are open clusters with high stellar densities as well, though these are typically very young and still associated with the star-forming clouds of their birth. There is no evidence for nebulosity surrounding GC01 or GC02 that would hint of current associations with such star-forming regions (such emission is usually brightest at $K_s$ and thus less affected by extinction). Likewise, emission from young clusters is dominated by hot, blue stars, while globulars contain much larger populations of red giant branch stars that are easier to detect in the near infrared. One further piece of circumstantial evidence is that the globular cluster distribution is highly concentrated towards the Galactic Center, and that these two clusters are sitting in the most enhanced region of that distribution (see Fig. 5). Granted this enhancement along with the high extinctions here it was likely that at least a few globulars would have escaped detection in earlier optical surveys, and it is possible that more may still be awaiting discovery.

Final confirmation of the nature of GC01 & GC02 will hinge on sensitive infrared spectroscopy to see whether the velocity dispersions of the stars indicate they are virially bound associations like other globulars. If confirmed, they will represent a notable expansion of the known set globulars, currently numbering 147 in the most recent catalog of Harris (1996). Given the large Galactic dust extinctions near the peak in the globular cluster distribution of the Milky Way (see Fig. 5) it is likely that more will be found in the 2MASS dataset.

## 4. Summary

The discoveries presented in this paper are serendipitous insofar as they were uncovered during normal data processing activities and not the result of any systematic search. They do emphasize the importance of visual study for identifying many interesting sources hiding within the 2MASS dataset. While galaxies like 2MASXI J0730080-220105 are automatically detected and available in the 2MASS Extended Source Catalog, proper interpretation of all extended sources the ZOA will rely on visual inspections because of the high stellar densities and source confusion here. Star clusters of all types are even more elusive since the extended source detection algorithms are optimized for galaxies; GC01 & GC02 have no corresponding entries in the Extended Source Catalog though their brightest member stars appear in the Point Source Catalog. It is unlikely that any practical automated search of the Point Source Catalog could reliably locate similarly faint associations in fields with such high source densities and extinctions. Since only a small fraction of the 2MASS survey has been inspected by eye during quality review and catalog production, our results are but a hint of the interesting objects waiting to be found within this huge dataset.

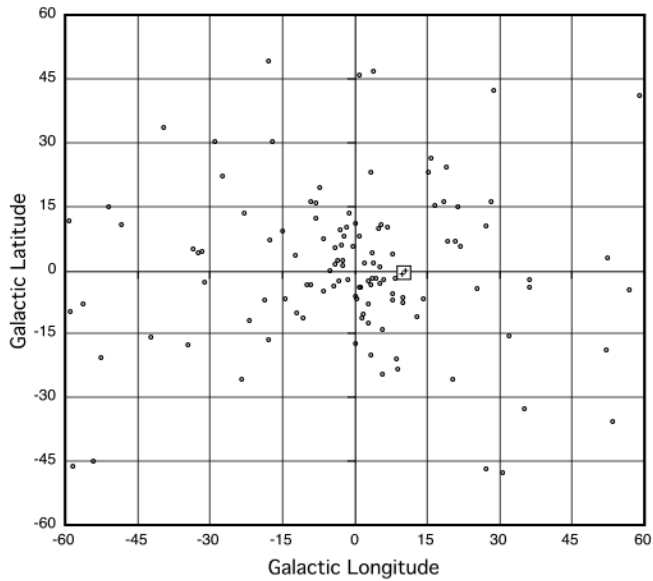

Fig. 5.—Galactic Globular Cluster Distribution. The locations of all prior known globular clusters (Harris 1996) with |l| < 60° & |b| < 60° are marked with circles. The locations of GC01 & GC02 are indicated by crosses near l = 10, b = 0, inside the small box.

RLH, THJ, JDK, & RMC acknowledge the support of the Jet Propulsion Laboratory, California Institute of Technology, which is operated under contract with NASA. This publication makes use of data products from 2MASS, which is a joint project of the University of Massachusetts and the Infrared Processing and Analysis Center/California Institute of Technology, funded by NASA and NSF. This research has made use of the NASA/IPAC Extragalactic Database (NED) which is operated by the Jet Propulsion Laboratory, California Institute of Technology, under contract with the National Aeronautics and Space Administration. We thank Ivan King for important feedback as well as the referee for constructive commentary.